\documentclass[prl,longbibliography,twocolumn,nopacs,preprintnumbers,notitlepage,amsmath,amssymb,superscriptaddress]{revtex4-2}

\usepackage{amsmath}
\usepackage{amssymb}
\usepackage{amsbsy}
\usepackage{graphicx}
\usepackage{xcolor}

\usepackage{mathrsfs}

\usepackage{enumerate}
\usepackage{mathtools}

\newcommand{\moy}[1]{\left\langle #1 \right\rangle}
\newcommand{\dd}[0]{\mathrm{d}}

\newcommand{\erf}[0]{\text{erf}}

\usepackage{soul}

\def\e{e}
\def\I{\mathrm{i}}

\newcommand{\ueta}{\{\eta_r \}}

\DeclareMathOperator{\erfc}{erfc}

\definecolor{darkblue}{rgb}{0,0,0.6}
\definecolor{darkred}{rgb}{0.6,0,0}
\usepackage[colorlinks=true,urlcolor=darkblue,citecolor=darkblue,linkcolor=darkred]{hyperref}

\def\rb{\bar\rho}
\def\rl{\rho_{\mathrm{L}}}

\def\qf{q^{\mathrm{(F)}}}
\def\pf{p^{\mathrm{(F)}}}

\begin{document}

\title{Semi-infinite simple exclusion process: from current fluctuations to target survival}

\author{Aur\'elien Grabsch}
\affiliation{Sorbonne Universit\'e, CNRS, Laboratoire de Physique Th\'eorique de la Mati\`ere Condens\'ee (LPTMC), 4 Place Jussieu, 75005 Paris, France}

\author{Hiroki Moriya}
\affiliation{Sorbonne Universit\'e, CNRS, Laboratoire de Physique Th\'eorique de la Mati\`ere Condens\'ee (LPTMC), 4 Place Jussieu, 75005 Paris, France}

\author{Kirone Mallick}
\affiliation{Institut de Physique Th\'eorique, CEA, CNRS, Universit\'e Paris–Saclay, F–91191 Gif-sur-Yvette cedex, France}

\author{Tomohiro Sasamoto}
\affiliation{Department of physics, Tokyo Institute of Technology, Tokyo 152-8551, Japan}

\author{Olivier B\'enichou}
\affiliation{Sorbonne Universit\'e, CNRS, Laboratoire de Physique Th\'eorique de la Mati\`ere Condens\'ee (LPTMC), 4 Place Jussieu, 75005 Paris, France}

\begin{abstract}
    The symmetric simple exclusion process (SEP), where diffusive particles cannot overtake each other, is a paradigmatic model of transport in the single-file geometry. In this model, the study of currents has attracted a lot of attention, but so far most results are restricted to two geometries: (i) a finite system between two reservoirs, which does not conserve the number of particles but reaches a nonequilibrium steady state, and (ii) an infinite system which conserves the number of particles but never reaches a steady state. Here, we obtain an expression for the full cumulant generating function of the integrated current in the important intermediate situation of a semi-infinite system connected to a reservoir, which does not conserve the number of particles and never reaches a steady state. This results from the determination of the full spatial structure of the correlations, which we infer to obey the very same closed equation recently obtained in the infinite geometry and argue to be exact. Besides their intrinsic interest, these results allow us to solve two open problems: the survival probability of a fixed target in the SEP, and the statistics of the number of particles injected by a localized source.
\end{abstract}

\maketitle

\let\oldaddcontentsline\addcontentsline
\renewcommand{\addcontentsline}[3]{}

\emph{Introduction.---} A key minimal model in statistical physics is the symmetric exclusion process (SEP)~\cite{Spitzer:1970,Spohn:1991,Derrida:2011,Chou:2011,Mallick:2015}. In this lattice gas model, particles perform symmetric random walks in continuous time and interact by hard-core exclusion, so that each particle attempts to hop with unit rate to an empty neighboring site. A basic observable which has received a lot of attention in the physics and mathematics literature is the total current $Q_t$ through a given point~\cite{Derrida:2004,Derrida:2009,Derrida:2009a,Krapivsky:2012,Zarfaty:2016,Mallick:2022,Mallick:2024}, defined as the total number of particles that have crossed this point from left to right, minus the number from right to left, up to time $t$. Interest in this quantity originates from its crucial role to quantify both out-of-equilibrium effects and thermal fluctuations.

Existing results concerning the current in the SEP can schematically be classified according to the nature of the geometry, either finite systems (between reservoirs or under periodic boundary conditions), or infinite systems. These situations correspond to different behaviors of the current which is stationary in the finite case, while never reaches a steady state in the infinite case.
Note that finite systems between reservoirs do not conserve the total number of particles, in contrast with periodic and infinite systems.
Relying on microscopic (integrable probability methods~\cite{Derrida:2009}) or macroscopic (fluctuating hydrodynamics and macroscopic fluctuation theory~\cite{Bertini:2015}) approaches, the statistical properties of the current $Q_t$ have been fully determined in the long time limit in these different geometries, both in finite systems, with reservoirs~\cite{Bodineau:2004,Derrida:2004,Bodineau:2007} or periodic boundary conditions~\cite{Appert:2008}, and in infinite systems~\cite{Derrida:2009,Derrida:2009a}.

On the other hand, the important situation of a semi-infinite geometry, describing a system connected to a single reservoir, has received far less attention up to now~\cite{Saha:2023}. In this intermediate situation, the number of particles is not conserved, and the system never reaches a steady state. The study of this geometry is of high relevance for two reasons. First, it gives access to the behavior of systems between reservoirs at intermediate time scales, before they reach a steady state~\footnote{For finite systems of length $L$, we expect that the time to establish a steady state is controlled by the time a particle injected on one side needs to diffuse to the other side, which grows as $L^2$. For times $1 \ll t \ll L^2$, near a reservoir, a finite system is expected to behave as a semi-infinite system. This is illustrated in more details in the Supplemental Material~\cite{SM}.}. Such transient behavior is out of reach of existing studies which focus on the long time regime~\cite{Bodineau:2004,Derrida:2004,Bodineau:2007}. Second, as a particular case, it allows one to describe the situation in the presence of an absorbing boundary. Such boundary conditions are known to play a key role in the important class of first-passage problems~\cite{Redner:2001,Bray:2013,Benichou:2014}, which find applications in fields as varied as random search strategies~\cite{Benichou:2011} or diffusion limited reactions~\cite{Rice:1985}.

While several works have been interested in exclusion processes in the semi-infinite geometry~\cite{Liggett:1975,Duhart:2018}, the only known results on the current in the SEP are very recent and concern the calculation of the full cumulant generating function (CGF) in the low density limit, and of the first three cumulants at arbitrary density~\cite{Saha:2023}.
Here, we determine the full CGF at arbitrary density. This result is obtained thanks to the characterization of the complete spatial structure of the current-density correlation functions. These correlations are determined by using the method recently introduced in~\cite{Grabsch:2022} which relies on a combination of microscopic and macroscopic calculations. In addition to these explicit expressions, our results have two important merits: (i) they show that the correlations satisfy the exact same integral equation as the one found in the infinite case~\cite{Grabsch:2022}, which further emphasises the universality of this equation; and (ii) they allow us to solve two open problems: the survival probability of a fixed target in the SEP, for which only perturbative expressions in the density have been obtained~\cite{Krapivsky:2014a}, and the statistics of the number of particles injected by a point source in the SEP, for which only the mean and variance are known~ \cite{Krapivsky:2012a,Krapivsky:2014c} in 1D.

\begin{figure}
    \centering
    \includegraphics[width=0.9\columnwidth]{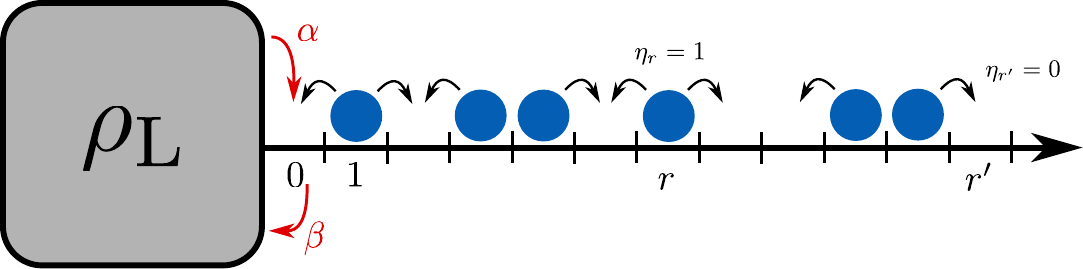}
    \caption{A symmetric exclusion process on the semi-infinite lattice $r \in \mathbb{N}$, connected on site $0$ to a reservoir at density $\rl$ to the left. The particles jump with unit rate in either direction to an empty site. The reservoir injects particles with rate $\alpha$ (if the site is empty), and absorbs particles with rate $\beta$. This dynamics describes a reservoir with density $\rl = \frac{\alpha}{\alpha + \beta}$.}
    \label{fig:SEP_half_inf}
\end{figure}

\emph{Model.---} We consider a SEP on a semi-infinite lattice $r \in \mathbb{N}$ (see Fig.~\ref{fig:SEP_half_inf}). Particles, present initially picked from an equilibrium distribution at density $\rb$, perform symmetric continuous-time random walks with unit jump rate, and with the hard-core constraint that there is at most one particle per site. 
This is described by occupation numbers $\eta_r(t)$ for $r \in \mathbb{N}$, defined such as $\eta_r(t) = 1$ if site $r$ is occupied at time $t$ and $\eta_r(t) = 0$ otherwise.
Site $0$ is connected to a reservoir which injects particles with rate $\alpha$ (if site $0$ is empty), and absorbs particles with rate $\beta$ (if site $0$ is occupied), so that the reservoir has density $\rl = \frac{\alpha}{\alpha+\beta}$. Due to these exchanges with the reservoir, the number of particles in the system is not conserved.
We are interested in the total number $Q_t$ of particles which have been exchanged with the reservoir (counted positively for injected particles and negatively for absorbed ones). The full statistics of $Q_t$ is described by the CGF, of parameter $\lambda$, whose long time behavior is encoded in the scaled CGF $\hat\psi(\lambda)$ defined as
\begin{equation}
\label{eq:DefCurrGenFct}
    \ln \moy{\e^{\lambda Q_t}} 
    \underset{t\to \infty}{\simeq}
    \sqrt{t} \: \hat\psi(\lambda)
    \:.
\end{equation}
The knowledge of $\hat\psi$ gives the long time behavior of the cumulants, for instance, $\moy{Q_t} \simeq \sqrt{t} \: \hat\psi'(0)$ and $\moy{Q_t^2}_c \equiv \moy{Q_t^2} - \moy{Q_t}^2 \simeq \sqrt{t} \: \hat\psi''(0)$.
The presence of a $\sqrt{t}$ in all the cumulants, instead of $t$, indicates that the system never reaches a steady state, which is general for single-file systems~\cite{Derrida:2009,Derrida:2009a}.
As we proceed to show, the key to characterize the statistics of $Q_t$ is to introduce and determine the current-density correlation profiles
\begin{equation}
\label{eq:DefGenProf}
    w_r(t) = 
    \frac{\moy{ \eta_r \: \e^{\lambda Q_t}} }{\moy{\e^{\lambda Q_t}}}
    = \sum_{n\geq 0} \moy{\eta_r \: Q_t^n}_c \frac{\lambda^n}{n!}
    \:,
\end{equation}
which generate all the connected correlation functions $\moy{Q_t^n \eta_r}_c$. For instance, $\moy{Q_t \eta_r}_c = \moy{Q_t \eta_r} - \moy{Q_t} \moy{\eta_r}$ measures the covariance between $Q_t$ and $\eta_r$. Additionally, the correlation profiles control
the time evolution of the cumulants (see SM~\cite{SM} for derivation)
\begin{equation}
    \label{eq:EvolCGFmicro}
    \frac{\dd }{\dd t} \ln \moy{\e^{\lambda Q_t}}
    = \alpha (\e^\lambda - 1)
    + (\e^{-\lambda}-1) [\beta + \alpha \e^{\lambda}] w_0(t)
    \:.
\end{equation}
These profiles are shown below to display a diffusive scaling with a scaling function $\Phi$ defined by
\begin{equation}
    \label{eq:ScalingProf}
    w_r(t)
    \underset{t\to \infty}{\simeq}
    \Phi \left( x = \frac{r}{\sqrt{t}} \right)
    \:,
\end{equation}
which indicates that the correlations are not stationary but propagate with time on a distance which grows as $\sqrt{t}$ away from the reservoir.

\emph{Results.---} We present here our main results. A sketch of the derivation is given below (see SM~\cite{SM} for details).
We show that the CGF of $Q_t$~\eqref{eq:DefCurrGenFct} at large times takes the form
\begin{equation}
  \label{eq:CumulGenFct}
    \hat\psi
     = - \frac{1}{2 \sqrt{\pi}} \:
    \mathrm{Li}_{\frac{3}{2}} \left[ - 4 \omega(1+\omega) \right]
    \:,
\end{equation}
where $\mathrm{Li}_s(x)$ is the polylogarithm of order $s$ and $\omega$ is the single-parameter combining $\rl$, $\rb$ and $\lambda$,
\begin{equation}
    \label{eq:Defomega}
  \omega = \rl (\e^{\lambda} - 1) + \rb (\e^{-\lambda}-1) + \rb \rl (\e^{\lambda}-1)(\e^{-\lambda}-1)
  \:,
\end{equation}
which appears both in finite~\cite{Bodineau:2004,Derrida:2004} and infinite geometries~\cite{Derrida:2009,Derrida:2009a,Imamura:2017,Imamura:2021}.

This result is obtained thanks to the determination of the full spatial structure of the current-density correlations, which are encoded in the scaling function $\Phi$~\eqref{eq:ScalingProf}. We indeed infer from the first $3$ orders in $\lambda$ that the rescaled derivative of $\Phi$,
\begin{equation}
  \label{eq:defOmega}
  \Omega(x) =  \hat\psi \frac{\Phi'(x)}{\Phi'(0)}
  \:,
\end{equation}
satisfies, at all orders in $\lambda$, the closed integral equation
\begin{equation}
  \label{eq:IntegEq}
    \Omega(x) + \int_0^\infty \Omega(z) \Omega(x+z) \dd z
    = \gamma \frac{\e^{- \frac{x^2}{4}}}{\sqrt{4 \pi}}
    \:,
\end{equation}
where $\gamma$ is a parameter determined by the boundary conditions
\begin{equation}
  \label{eq:ProfAtZero}
    \Phi(0) = \frac{\alpha}{\alpha + \beta \e^{-\lambda}}
    = \frac{\rl \e^\lambda}{1 + (\e^\lambda - 1) \rl}
    \:,
    \quad
    \Phi(+\infty) = \bar\rho
\end{equation}
\begin{equation}
  \label{eq:DerProfZero}
    \Phi'(0) =  \frac{\hat\psi}{2} \left( \frac{1}{\e^{-\lambda}-1} + \Phi(0) \right)
    \:.
\end{equation}
We argue below that the integral equation~(\ref{eq:IntegEq}) is exact. It
can be solved explicitly for $\Omega$, from which the spatial structure of the correlations, encoded in $\Phi$, are deduced (see Eq.~\eqref{eq:SolIntegEq} and consequences below).
For instance, the lowest orders in $\lambda$, defined by $\Phi(x) = \Phi_0(x) + \lambda \Phi_1(x) + \cdots$, are given by (see~\cite{SM} for further orders in $\lambda$)
\begin{equation}
    \label{eq:Phi0}
    \Phi_0(x) = \rl \erfc \left( \frac{x}{2} \right)
    + \rb \: \erf \left( \frac{x}{2} \right)
    \:,
\end{equation}
\begin{equation}
    \label{eq:Phi1}
    \Phi_1(x) = (\rb^2 + \rl (1-2\rb)) \erfc \left( \frac{x}{2} \right)
    - (\rb - \rl)^2 \erfc \left( \frac{x}{2 \sqrt{2}} \right)
    \:.
\end{equation}

Besides their intrinsic interest, these results allow us (i) to further demonstrate the potential of universality of the equation obtained in~\cite{Grabsch:2022} in the infinite geometry, (ii) to illustrate the generality of the ``physical'' boundary conditions recently derived in~\cite{Grabsch:2024} in an infinite geometry, and (iii) to solve the open problem of the survival probability of a fixed target in the SEP.

\emph{A universal equation for the correlations.---} We can show that the integral equation~(\ref{eq:IntegEq}) is
equivalent to the ones obtained in~\cite{Grabsch:2022,Grabsch:2023}, which in turn provides the solution of~(\ref{eq:IntegEq}). For
this, we introduce
\begin{equation}
  \label{eq:DefOmegaPM}
  \Omega_\pm(x) = \Omega(\pm x)
  \:,
  \quad \text{for} \quad
  x \gtrless 0
  \:.
\end{equation}
We can rewrite~(\ref{eq:IntegEq}) as two equations for $\Omega_\pm$, for $x \gtrless 0$,
\begin{equation}
  \label{eq:CoupledWHeqs}
    \Omega_\pm(x) + \int_0^\infty \Omega_\mp(\mp z) \Omega_\pm(x\pm z) \dd z
    = K(x)
      \:,
\end{equation}
where we have introduced
\begin{equation}
  \label{eq:DefKernel}
  K(x) = \gamma 
  \frac{\e^{- \frac{x^2}{4}}}{\sqrt{4 \pi}}
  \quad \text{for} \quad x \in \mathbb{R}
  \:.
\end{equation}
These are exactly the equations written
in~\cite{Grabsch:2022,Grabsch:2023} for the infinite system. The
mapping~(\ref{eq:DefOmegaPM}) is indeed valid since the solution
of~(\ref{eq:CoupledWHeqs}) given in~~\cite{Grabsch:2022,Grabsch:2023} is
symmetric for a symmetric kernel $K$. This gives the solution
\begin{multline}
  \label{eq:SolIntegEq}
  \int_0^\infty \Omega(x) \e^{\I k x} \dd x
  =   \\
  \exp \left[
    \frac{1}{2\pi} \int_0^\infty \dd x \: \e^{\I k x}
    \int_{-\infty}^{+\infty} \dd u \: \e^{-\I u x} \:
    \ln (1+\hat{K}(u))
  \right]-1
  \:,
\end{multline}
where
\begin{equation}
  \hat{K}(u) = \int_{-\infty}^\infty K(x) \e^{\I k x} \dd x
  = \gamma \: \e^{- k^2}
  \:.
\end{equation}
In particular, we obtain the CGF from $\Omega(0) = \hat\psi = -\frac{1}{2\sqrt{\pi}} \mathrm{Li}_{\frac{3}{2}}(-\gamma)$, deduced from~\eqref{eq:SolIntegEq} by setting $k = \I s$ and letting $s \to \infty$. The expression of $\gamma$ in terms of $\rl$, $\rb$ and $\lambda$ can be obtained by combining the explicit solution~\eqref{eq:SolIntegEq} with the boundary conditions~(\ref{eq:ProfAtZero},\ref{eq:DerProfZero}), see SM~\cite{SM}. This leads to
\begin{equation}
    \label{eq:ExprGamma}
    \gamma = 4 \omega (1+\omega)
    \:,
\end{equation}
with $\omega$ given by~\eqref{eq:Defomega}, and thus to the CGF~\eqref{eq:CumulGenFct}.

We stress that, in addition to the CGF, our approach provides the full spatial structure of the correlations between the current $Q_t$ and the density of particles, encoded in $\Phi$. It can be computed by integrating Eq.~\eqref{eq:defOmega} with the boundary condition at infinity~\eqref{eq:ProfAtZero} and at the origin~(\ref{eq:ProfAtZero},\ref{eq:DerProfZero}). Explicitly, this procedure provides for instance the expressions~(\ref{eq:Phi0},\ref{eq:Phi1}).

A key aspect of our results is that the closed equations~\eqref{eq:CoupledWHeqs} satisfied by the (rescaled derivative of the) correlation profile $\Phi$ in the semi-infinite geometry is exactly the \textit{same} as the one recently unveiled in the infinite case~\cite{Grabsch:2022}. This shows that this equation, which has been shown to hold in a variety of situations for infinite systems~\cite{Grabsch:2022,Grabsch:2023,Grabsch:2024} ---out-of-equilibrium cases, other observables than the current, other single-file models than the SEP--- also applies to semi-infinite systems. The robustness of the equation with respect to the geometry of the system further demonstrates the potential of universality of this closed equation.

\emph{Physical form of the boundary conditions.---} An interesting by-product of our approach are the boundary conditions~(\ref{eq:ProfAtZero},\ref{eq:DerProfZero}). Indeed, it has recently been shown that, for the infinite geometry, boundary conditions associated with the current through the origin in a single-file system can be written in a physical form in terms of the chemical potential $\mu(\rho)$ and the collective diffusion coefficient $D(\rho)$~\cite{Grabsch:2024,Grabsch:2024a}, as
\begin{equation}
    \label{eq:RelChemPot}
    \mu(\Phi(0)) - \mu(\rl) = \lambda
    \:,
\end{equation}
\begin{equation}
    \label{eq:RelPsiChemPot}
    \hat\psi = - 2 \left. \partial_x \mu(\Phi) \right|_{x=0}
    \int_{\rl}^{\Phi(0)} D(r)\dd r
    \:,
\end{equation}
where for the SEP,
\begin{equation}
    \mu(\rho) = - \ln \left( \frac{1}{\rho} - 1 \right)
    \:,
    \quad
    D(\rho) = 1
    \:.
\end{equation}
The advantage of such a physical reformulation is that it applies to general diffusive single-file systems~\cite{Grabsch:2024,Grabsch:2024a}.
Our results on the boundary conditions~(\ref{eq:ProfAtZero},\ref{eq:DerProfZero}) can be recast into the exact same expressions~(\ref{eq:RelChemPot},\ref{eq:RelPsiChemPot}), demonstrating that the physical relations obtained in~\cite{Grabsch:2024,Grabsch:2024a} still hold in the semi-infinite SEP. In turn, this suggests that the boundary conditions~(\ref{eq:RelChemPot},\ref{eq:RelPsiChemPot}) hold for any single-file system, beyond the SEP, in the semi-infinite geometry.

\emph{Survival probability of a fixed target in the SEP.---} As an application of our results, we show that they allow us to solve the problem of the survival probability of a fixed target in the SEP.
It is defined as the probability that no particle (representing for instance diffusive reactants), initially uniformly distributed with density $\rb$, has reached a fixed target (the other reactive species) up to time $T$.
In the case of independent particles, this constitutes a classical problem of chemical physics which has received a lot of attention~\cite{Burlatsky:1987,Blumen:1984,Szabo:1988,Blumen:1986,Benichou:2000}. Despite its importance, its extension to the case of interacting particle systems has essentially been left aside so far. The only contributions to the problem in the SEP have been performed in the important Ref.~\cite{Krapivsky:2014a} (see also~\cite{Agranov:2018} for a related problem). In the 1D case, the survival probability has been determined to second order in the density of particles $\rb$ (low density limit). However, the determination of the survival probability at arbitrary density in the SEP constitutes an open problem.

We first remark that the fixed target (placed at the origin) corresponds to a reservoir with density $\rl = 0$, because it can only absorb particles but not inject any. Next, the target has survived up to time $T$ if and only if no particle has crossed the origin up to time $T$. In other words,
\begin{equation}
  S(T) \equiv \mathbb{P}(\text{Surv. up to } T) = \mathbb{P}(Q_T = 0)
  \:,
\end{equation}
which in turn can be obtained from the results presented above. Indeed, the distribution of $Q_T$ can be deduced from the CGF $\hat\psi$~\eqref{eq:CumulGenFct} by an inverse Laplace transform, which for large $T$ reduces to a Legendre transform. This gives,
\begin{equation}
  \mathbb{P}(Q_T = q \sqrt{T}) \underset{T \to \infty}{\simeq} \e^{- \sqrt{T} \phi(q)}
  \:,
\end{equation}
where
\begin{equation}
    \phi(q) = - \hat\psi(\lambda^\star(q)) 
    - q \lambda^\star(q)
  \:,
  \quad \text{and} \quad
  \hat\psi'(\lambda^\star(q)) = q
  \:.
\end{equation}
Since for $\rl =0$, $\hat\psi'(\lambda) \propto \e^{-\lambda}$, the solution of $\hat\psi'(\lambda^\star) = 0$ corresponds to the limit $\lambda \to \infty$. Therefore, the survival probability reads
\begin{equation}
    \label{eq:SurvProb}
    S(T)  \underset{T \to \infty}{\simeq} \exp \left[
      -\sqrt{\frac{T}{4 \pi}} \: \mathrm{Li}_{\frac{3}{2}}(4\rb(1-\rb))
    \right]
    \:.
\end{equation}
The first two orders in $\rb$ coincide with those computed in~\cite{Krapivsky:2014a}. Equation~\eqref{eq:SurvProb} finally provides the solution for arbitrary density $\rb \leq \frac{1}{2}$.
Note that, for $\rb > \frac{1}{2}$, the CGF $\hat\psi$ has a non-analyticity for $\lambda = \log \frac{2\rb}{2 \rb - 1} > 0$, and the above procedure cannot be applied. Additionally, for $\rb \to 1$, the survival probability is fully determined by the time needed by the first particle to jump into the reservoir, which is exponentially distributed with rate $\beta$. Hence, we expect that the non-analytic behavior of $\hat\psi$ indicates a phase transition to a regime of exponentially distributed survival times $\rb > \frac{1}{2}$. The precise determination of this regime requires further investigation.

\emph{SEP with a localized source.---} A second example illustrating the key role of our results is provided by the problem of the number of particles injected by a point source in the SEP. This problem, introduced in~\cite{Krapivsky:2012a,Krapivsky:2014c}, consists in studying a SEP initially empty, connected to a source which injects particles on a given site at a rate $\alpha$, if the site is empty. This model offers a minimal description of a growth process in an initially empty medium, with hardcore interactions. It is also relevant in fields as varied as monomer-monomer catalysis~\cite{Krapivsky:1992,Krapivsky:1996}, the voter model~\cite{Mobilia:2003}, or the spreading of thin wetting films~\cite{Popescu:2012}.

The SEP with a localized source with fast injection rate $\alpha \to \infty$ actually appears as the particular case $\beta = 0$ (thus $\rl = 1$) and $\rb = 0$ of the model considered here. The CGF of the number $N_t$ of particles injected by the source at time $t$ is therefore given by~\eqref{eq:CumulGenFct} with $\rl = 1$ and $\rb = 0$,
\begin{equation}
    \frac{1}{\sqrt{t}} \ln \moy{ \e^{\lambda N_t}}
    \underset{t \to \infty}{\simeq} - \frac{1}{2\sqrt{\pi}}
    \mathrm{Li}_{\frac{3}{2}}[ -4 \e^\lambda(\e^\lambda - 1) ]
    \:.
\end{equation}
This expression, which reproduces the first two cumulants of $N_t$ (see SM~\cite{SM} for details), which were the only previously known results~\cite{Krapivsky:2012a,Krapivsky:2014c}, provides the full CGF of the number of particles injected by the source.

\emph{Mains steps of the derivation.---} We now sketch the main steps that led to the closed equation~\eqref{eq:IntegEq} and the boundary conditions~(\ref{eq:ProfAtZero},\ref{eq:DerProfZero}). The details of the derivation are given in SM~\cite{SM}.

First, concerning the boundary conditions, we start from a microscopic description of the system, in terms of the master equation for the occupation numbers $\{ \eta_r \}_{r \in \mathbb{N}}$. From this master equation, we deduce the time evolution of $\moy{\e^{\lambda Q_t}}$ and $\moy{\eta_0 \: \e^{\lambda Q_t}}$, and thus the equations satisfied by $\ln \moy{\e^{\lambda Q_t}}$~\eqref{eq:EvolCGFmicro} and $w_0(t)$. Inserting in these equations the long time behaviors~(\ref{eq:DefCurrGenFct},\ref{eq:ScalingProf}), we finally get the boundary conditions~(\ref{eq:ProfAtZero},\ref{eq:DerProfZero}).

Second, for the integral equation~\eqref{eq:IntegEq}, we follow a different approach. We start from a macroscopic description of the system in terms of a (stochastic) density field $\rho(x,t)$, following the approach of fluctuating hydrodynamics~\cite{Spohn:1983}. One can then use a path integral formulation, from which the long time behavior of the system can be obtained through the minimization of an action. This is the formalism of macroscopic fluctuation theory (MFT), which has been applied to various systems and observables~\cite{Bertini:2015}. In the case of the current $Q_T$ in the half-infinite SEP, the CGF can be obtained from the solution of the MFT equations~\cite{Derrida:2009a,Bertini:2015,Saha:2023},
\begin{align}
  \label{eq:MFT_q}
  \partial_t q &= \partial_x^2 q - 2\partial_x[q(1-q)\partial_x p]
  \:,
  \\
  \label{eq:MFT_p}
  \partial_t p &= - \partial_x^2 p -(1-2q)(\partial_x p)^2
  \:,
\end{align}
\begin{equation}
    \label{eq:InitFin}
    p(x,T) = \lambda
    \:,
    \quad
    p(x,0) = \lambda +  \int_{\rb}^{q(x,0)} \frac{\dd r}{r(1-r)}
    \:,
\end{equation}
\begin{equation}
    \label{eq:BoundCondMFT}
    q(0,t) = \rl
    \:,
    \quad
    p(0,t) = 0
    \:.
\end{equation}
In these equations, $q(x,t)$ represents the optimal fluctuation of the stochastic density $\rho(x,t)$ that realises the current $Q_{t=T}$. The field $p(x,t)$ is a Lagrange multiplier that enforces the local conservation of the number of particles at all points in space and time. The boundary conditions~\eqref{eq:BoundCondMFT} implement the connection of a reservoir at density $\rl$ to site $0$. The correlation profile~(\ref{eq:DefGenProf},\ref{eq:ScalingProf}) can be deduced from the solution of the MFT equations as $\Phi(x) = q(x,T)$~\cite{Poncet:2021,Grabsch:2022,Grabsch:2023}. 

In the case of an infinite system, closely related equations have been first solved perturbatively, and it was  inferred that the solution obeys the closed integral equations~\eqref{eq:CoupledWHeqs}, with the definition of $\Omega$~\eqref{eq:defOmega}, which was solved nonperturbatively, leading to the full spatial structure of the correlations~\cite{Grabsch:2022,Grabsch:2023}.
Later, the closed integral equations~\eqref{eq:CoupledWHeqs} were proved~\cite{Mallick:2022} using the integrability of the MFT equations~(\ref{eq:MFT_q},\ref{eq:MFT_p}) and the inverse scattering technique~\cite{Ablowitz:1974,Bettelheim:2022,Bettelheim:2022a,Mallick:2022,Krajenbrink:2021,Krajenbrink:2022,Krajenbrink:2022a,Mallick:2024}.
In the case of a semi-infinite system considered here, this latter approach cannot be straightforwardly applied~\footnote{In particular because the inverse scattering approach uses a mapping which involves the derivatives of both $q$ and $p$~\cite{Mallick:2022}, but we only have boundary conditions of the values of $q$ and $p$ at the origin~\eqref{eq:BoundCondMFT}, not their derivatives.}, so we rely on the perturbative approach (see SM~\cite{SM} for details). We have computed the solution of the MFT equations at final time $q(x,1) = \Phi(x)$ up to order $3$ in $\lambda$. From this solution, we build the function $\Omega$ as defined in~\eqref{eq:defOmega}. We then look for an integral equation similar to the ones found in the infinite case~\eqref{eq:CoupledWHeqs}, with the main difference that here we have only the domain $x>0$ since the system is semi-infinite. Plugging the perturbative expression of $\Omega$ in the l.h.s. of~\eqref{eq:IntegEq}, many terms cancel out, and there only remains the r.h.s.~\eqref{eq:IntegEq}, with a constant $\gamma$ found to be $4 \omega(1+\omega)$, at least up to order $3$ in $\lambda$. We then infer, based on the similarity with~\cite{Grabsch:2023,Grabsch:2024} and numerical evidence provided in SM~\cite{SM}, that this equation holds at all orders in $\lambda$.
We check that this expression is consistent with the microscopic boundary conditions~(\ref{eq:DefGenProf},\ref{eq:ScalingProf}), as shown with the procedure given before~\eqref{eq:ExprGamma}. We therefore claim that the closed equation~\eqref{eq:IntegEq} is exact, and thus the CGF~\eqref{eq:CumulGenFct} also.

\emph{Conclusion.---} We have obtained an expression for the full CGF of the integrated current in the SEP on a semi-infinite line, which constitutes a benchmark geometry to study the transient regime of systems connected to reservoirs, and access first-passage properties. Besides its intrinsic interest in statistical physics, this result allowed us to solve two open problems: (i) the key question in chemical physics of the survival probability of a fixed target in the presence of hardcore interacting random searchers; and (ii) the statistics of the number of particles injected by a localized source in the SEP, which provides a minimal model of the spreading of thin wetting films.

All these results are obtained thanks to the determination of the correlations between the current and the density of particles, which in turn provides the full spatial structure of the system. A fundamental point is that the closed equation~\eqref{eq:IntegEq} satisfied by these correlations is exactly the same as the one recently discovered in the case of an infinite geometry~\cite{Grabsch:2022}.
We have inferred this key equation from the first orders in $\lambda$ of the correlations, and argued that it holds for arbitrary $\lambda$. The robustness of this closed equation with respect to the geometry of the system further demonstrates its key role in the field of interacting particle systems.

\emph{Acknowledgements.---} The work of KM has been supported by the project RETENU ANR-20-CE40-0005-01 of the French National Research Agency (ANR). TS thanks a hospitality during his stays at the Isaac Newton Institute of Mathematical Sciences and  at Institut des Hautes \'Etudes Scientifiques where part of this work was done. The work of TS has been supported by JSPS KAKENHI Grants No. JP21H04432, No. JP22H01143. OB warmly thanks P. Krapivsky for stimulating discussions about the problem of a localized source in the SEP.


%

\clearpage
\widetext

\let\addcontentsline\oldaddcontentsline

\setcounter{equation}{0}
\setcounter{figure}{0}
\setcounter{table}{0}
\setcounter{page}{1}
\makeatletter
\renewcommand{\theequation}{S\arabic{equation}}
\renewcommand{\thefigure}{S\arabic{figure}}
\renewcommand{\bibnumfmt}[1]{[S#1]}
\renewcommand{\citenumfont}[1]{S#1}

\setcounter{equation}{0}
\makeatletter

\renewcommand{\theequation}{S\arabic{equation}}
\renewcommand{\thefigure}{S\arabic{figure}}

\renewcommand{\bibnumfmt}[1]{[S#1]}
\renewcommand{\citenumfont}[1]{S#1}

\setcounter{secnumdepth}{3}

\begin{center}
  \begin{large}

    \textbf{
     Supplemental Material for\texorpdfstring{\\}{} Semi-infinite simple exclusion process:\texorpdfstring{\\}{}  from current fluctuations to target survival
   }
  \end{large}
   \bigskip

    Aurélien Grabsch, Hiroki Moriya, Kirone Mallick, Tomohiro Sasamoto and Olivier Bénichou
\end{center}

\tableofcontents

\section{Microscopic equations}

In this Section, we derive microscopic equations following the approach of Refs.~\cite{Poncet:2021SM,Grabsch:2022SM,Grabsch:2023SM}.

\subsection{The master equation}
\label{sec:MastEq}

We describe a configuration of the SEP at time $t$ on the positive lattice by a set of occupation numbers $\{ \eta_r(t) \}_{r \in \mathbb{N}}$.
The system can be described in terms of a master equation for the probability to observe a given configuration at time $t$,
\begin{multline}
    \label{eq:MasterEqSM}
    \partial_t P_t(\ueta) =
    \sum_{r \geq 0} \left[ P_t(\ueta^{r,+}) - P_t(\ueta) \right]
    + \alpha \left[\eta_0 \: P_t(\ueta^{(0)}) - (1-\eta_0) P_t(\ueta) \right]
    \\
    + \beta  \left[ (1-\eta_0) P_t(\ueta^{(0)}) - \eta_0 \: P_t(\ueta) \right]
    \:,
\end{multline}
where $\ueta^{r,+}$ corresponds to the configuration $\ueta$ in which the occupations $\eta_r$ and $\eta_{r+1}$ have been exchanged, $\ueta^{(0)}$ is the configuration in $\ueta$ with $\eta_0$ replaced by $1-\eta_0$. The first term corresponds to the hopping of the particles on the line, the second term the injection of particles on site $0$ with rate $\alpha$ (if the site was free before the injection), and the last term absorption of particles with rate $\beta$ (if the site was occupied before the absorption). In principle, the terms with $\eta_{r+1} = \eta_r$ should be removed from the first term due to the exclusion rule. However, since in this case $\ueta^{r,+} = \ueta$, these terms cancel out automatically.

\subsection{Evolution equations}

From the master equation~\eqref{eq:MasterEqSM}, we can compute the evolution of $\moy{\e^{\lambda Q_t}}$, as
\begin{equation}
    \label{eq:EvolMGFSM}
    \partial_t \moy{\e^{\lambda Q_t}}
    = \sum_{\ueta} \e^{\lambda Q_t[\ueta]} \partial_t P_t(\ueta)
    = \alpha (\e^{\lambda} - 1 ) \moy{ (1-\eta_0) \e^{\lambda Q_t} }
    + \beta ( \e^{-\lambda} -1) \moy{\eta_0 \e^{\lambda Q_t}}
    \:.
\end{equation}
Similarly, we obtain the evolution equation
\begin{equation}
    \label{eq:EvolProf00SM}
    \partial_t \moy{\eta_0 \: \e^{\lambda Q_t}}
    = \moy{(\eta_1 - \eta_0) \e^{\lambda Q_t}}
    + \alpha \e^{\lambda} \moy{(1-\eta_0) \e^{\lambda Q_t}}
    - \beta \moy{\eta_0 \e^{\lambda Q_t}}
    \:.
\end{equation}
From~\eqref{eq:EvolMGFSM}, we deduce the equation for the time evolution of the CGF given in the main text,
\begin{equation}
    \label{eq:TimeEvolCGFSM}
    \partial_t \ln \moy{\e^{\lambda Q_t}}
    = \alpha (\e^{\lambda} - 1)
     + [ \beta (\e^{-\lambda} - 1) - \alpha (\e^{\lambda}-1) ] w_0(t)
     \:,
\end{equation}
with
\begin{equation}
    w_r(t) = \frac{\moy{\eta_r \: \e^{\lambda Q_t}}}{\moy{\e^{\lambda Q_t}}}
    = \sum_{n \geq 0} \frac{\lambda^n}{n!} \moy{\eta_r \: Q_t^n}_c
\end{equation}
the generating function of the connected correlations between the density on site $r$ and the current $Q_t$. For instance, for $n=1$, this gives $\moy{Q_t \: \eta_r}_c = \moy{Q_t \: \eta_r} - \moy{Q_t} \moy{\eta_r}$, which is the covariance between $Q_t$ and $\eta_r$. Additionally, from~\eqref{eq:EvolProf00SM}, we obtain the time evolution of $w_0(t)$,
\begin{equation}
    \partial_t w_0(t) =
    w_1(t) - w_0(t)
    + \alpha \e^\lambda
    - (\alpha \e^{\lambda} + \beta) w_0(t)
    - w_0(t) \partial_t \ln \moy{\e^{\lambda Q_t}}
    \:.
\end{equation}
To study the long time behaviour, it is convenient to rewrite this equation in terms of~\eqref{eq:TimeEvolCGFSM} as
\begin{equation}
    \label{eq:EvolProf0SM}
    \partial_t w_0(t) =
    w_1(t) - w_0(t)
    - \frac{\partial_t \ln \moy{\e^{\lambda Q_t}}}{\e^{-\lambda} - 1}
    - w_0(t) \partial_t \ln \moy{\e^{\lambda Q_t}}
    \:.
\end{equation}
Equations~(\ref{eq:TimeEvolCGFSM},\ref{eq:EvolProf0SM}) are the starting point to derive the boundary conditions in the long time limit.

\subsection{Long time behaviour}

In the long time limit, the cumulant generating function and the correlation profiles obey the scaling forms (see section Macroscopic Fluctuation Theory below for a proof),
\begin{equation}
    \label{eq:ScalingFromSM}
    \ln \moy{\e^{\lambda Q_t}}
    \underset{t \to \infty}{\simeq}
    \sqrt{t} \: \hat\psi(\lambda)
    \:,
    \quad
    w_r(t) 
    \underset{t \to \infty}{\simeq}
    \Phi \left( x = \frac{r}{\sqrt{t}} \right)
    \:.
\end{equation}
Using these scaling forms in the equations for the CGF~\eqref{eq:TimeEvolCGFSM}, we obtain for large $t$, at leading order
\begin{equation}
    0 = \alpha (\e^\lambda - 1)
    + [\beta (\e^{-\lambda} - 1) - \alpha (\e^\lambda - 1)] \Phi(0)
    \:.
\end{equation}
This yields the value of the correlation profile at $0$,
\begin{equation}
    \label{eq:ValAtZeroSM}
    \Phi(0) = 
    \frac{\alpha }{\beta \e^{-\lambda} + \alpha}
    = \frac{\rl \e^\lambda}{1 + \rl (\e^{\lambda} -1)}
    \:,
    \quad
    \rl = \frac{\alpha}{\alpha + \beta}
    \:.
\end{equation}
This is the first boundary condition given in the main text, which coincides with the first two orders in $\rl$ given in~\cite{Derrida:2019bSM}.

Similarly, plugging the scaling forms~\eqref{eq:ScalingFromSM}  into the evolution equation for $w_0$~\eqref{eq:EvolProf0SM}, we get at leading order,
\begin{equation}
    \label{eq:BoundCondMicroSM}
    0 = \Phi'(0) - \frac{\hat\psi}{2} \left( \frac{1}{\e^{-\lambda} - 1} + \Phi(0) \right)
    \:.
\end{equation}
This is the second boundary condition given in the main text. Remarkably, it is identical to the one derived in the infinite case~\cite{Poncet:2021SM,Grabsch:2022SM,Grabsch:2023SM}.

\subsection{Finite systems: intermediate regime vs steady state}

The semi-infinite system considered in this manuscript can be used to describe the dynamics of a finite size system between two reservoirs at intermediate times: long time compared to the waiting time of an individual particle (to justify a macroscopic treatment), but short compared to the time needed by the system to reach a steady state. To illustrate this point, let us consider a system of $L$ sites, with occupations $\{ \eta_1(t) , \ldots, \eta_{L}(t) \}$. We consider that site $1$ is connected to a reservoir which injects particles with rate $\alpha$ and removes particles with rate $\beta$. Similarly, the right reservoir is connected to another reservoir which injects particles at rate $\gamma$ and absorbs them at rate $\delta$. We introduce the densities of the reservoirs, $\rl = \alpha/(\alpha + \beta)$ and $\rho_{\mathrm{R}} = \gamma/(\gamma + \delta)$. Initially, we consider that each site of the lattice is occupied with probability $\rb = \rho_{\mathrm{R}}$, so that the system is initially at equilibrium with the right reservoir.

\begin{figure}
    \centering
    \includegraphics[width=0.6\textwidth]{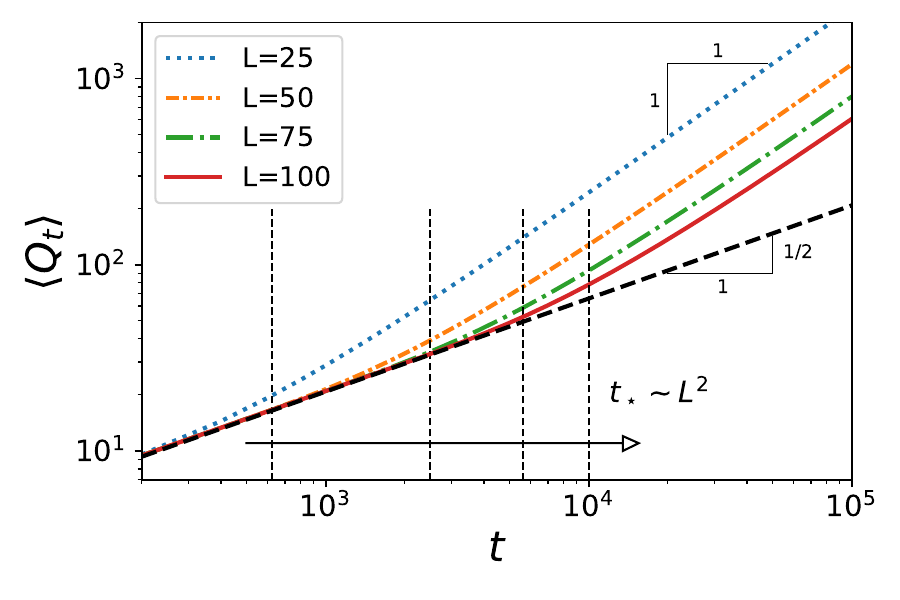}
    \caption{Mean integrated current $\moy{Q_t}$ for a finite system of length $L$, connected to reservoirs at densities $\rl = 0.75$ and $\rho_{\mathrm{R}} = \rb = 0.16$, obtained from~(\ref{eq:EvolMoyQtFinite}-\ref{eq:EvolMoyOccBoundFinite}), for different system sizes. In a first regime, the mean current follows the predicted behavior for a semi-infinite system (black dashed line). After a crossover time $t_\star$, which scales as $L^2$, the mean current grows linearly with time, indicating that the system has reached a steady state.}
    \label{fig:CrossoverFinite}
\end{figure}

Let us for simplicity consider the mean integrated current $\moy{Q_t}$, and the associated mean occupations $\moy{\eta_r(t)}$. The time evolution of these quantities can be written explicitly by using a formalism based on a master equation, as in Section~\ref{sec:MastEq}. We obtain,
\begin{equation}
    \label{eq:EvolMoyQtFinite}
    \partial_t \moy{Q_t} = \alpha - (\alpha + \beta) \moy{\eta_1(t)}
    \:,
\end{equation}
\begin{equation}
    \label{eq:EvolMoyOccFinite}
    \partial_t \moy{\eta_r(t)} =
    \moy{\eta_{r+1}(t)} - 2 \moy{\eta_r(t)} + \moy{\eta_{r-1}(t)}
    \:,
    \quad
    2 \leq r \leq L-1
    \:,
\end{equation}
\begin{equation}
    \label{eq:EvolMoyOccBoundFinite}
    \partial_t \moy{\eta_1(t)} = \moy{\eta_2(t)} + \alpha - (1+\alpha + \beta) \moy{\eta_1(t)}
    \:,
    \quad
    \partial_t \moy{\eta_L(t)} = \moy{\eta_{L-1}(t)} + \gamma - (1+\gamma + \delta) \moy{\eta_L(t)}
    \:.
\end{equation}
These equations can be solved, either analytically or numerically, to yield the mean integrated current $\moy{Q_t}$ as a function of time. The result of a numerical resolution is shown in Fig.~\ref{fig:CrossoverFinite}. It shows that there exists a regime in time, before a crossover time $t_\star$, in which the mean current follows the behaviour predicted for a semi-infinite system,
\begin{equation}
    \moy{Q_t} \simeq 2\frac{\rl - \rb}{\sqrt{\pi}} \sqrt{t}
    \:.
\end{equation}
After this crossover time, the mean current grows linearly with time, indicating that the system has reached a steady state. The time $t_\star$ needed to reach the steady state grows as $L^2$, which is expected due to the diffusive nature of the model. We explicitly showed here this transition of the mean integrated current, since it can be computed from the microscopic model, but the same is expected to hold for the fluctuations and higher order cumulants. Indeed, before the time $t_\star$, no particle could have diffused through the entire lattice, and thus the finite nature of the system is irrelevant.

Therefore, the results obtained in this manuscript for a semi-infinite system correctly describe the behavior of a finite system near a boundary in the regime $1 \ll t \ll L^2$.

\section{Macroscopic description: Macroscopic Fluctuation Theory}

The boundary equations have been conveniently derived from the microscopic description of the system. Following the same approach for the bulk equation satisfied by $w_r(t)$ is possible, but it yields an infinite hierarchy of equations which cannot be solved. Therefore, we turn directly to a macroscopic description of the system.

\subsection{Macroscopic Fluctuation Theory (MFT)}

The macroscopic fluctuation theory relies on a coarse-grained description of the SEP in terms of a density field $\rho(x,t)$ which obeys a stochastic diffusion equation~\cite{Spohn:1983SM}
\begin{equation}
    \label{eq:StochDiffSM}
    \partial_t \rho = \partial_x \left[
    D(\rho) \partial_x \rho
    + \sqrt{\sigma(\rho)} \xi
    \right]
    \:,
    \quad
    \text{with} \quad
    D(\rho) = 1   
    \quad
    \text{and} \quad
    \sigma(\rho) = 2 \rho(1-\rho)
    \:,
\end{equation}
where $\xi$ is a Gaussian white noise in space and time, with $\moy{\xi(x,t) \xi(x',t')} = \delta(x-x') \delta(t-t')$.
Note that this formalism extends to any 1D diffusive system, by adapting the two transport coefficients ---the collective diffusion coefficient $D(\rho)$ and the mobility $\sigma(\rho)$--- to the system under consideration.

This stochastic hydrodynamics formulation can be rewritten in terms of an action for the time evolution of the density~\cite{Bertini:2015SM}. In the case of an infinite system, this was used to write the cumulant generating function of the integrated current $Q_t$ as~\cite{Derrida:2009aSM},
\begin{equation}
    \label{eq:MGFfromMFTSM}
    \moy{\e^{\lambda Q_T}}
    =
    \int \mathcal{D} \rho(x,t) \mathcal{D} H(x,t)
    \int \mathcal{D} \rho(x,0) \:
    \e^{\lambda Q_T[\rho] - S[\rho,H] - F[\rho(x,0)]}
    \:,
\end{equation}
where $S$ is the MFT action ($H$ is a Lagrange multiplier that enforces the conservation of particles at every point in space and time)
\begin{equation}
    S[\rho,H] =
    \int_{-\infty}^\infty \dd x \int_0^T \dd t \left[
    H \partial_t \rho
    + D(\rho) \partial_x \rho \partial_x H
    - \frac{\sigma(\rho)}{2} (\partial_x H)^2
    \right]
    \:,
\end{equation}
$F$ gives the distribution of the initial condition $\rho(x,0)$ picked from an equilibrium density $\rb$,
\begin{equation}
    F[\rho(x,0)] = \int_{-\infty}^\infty \dd x
    \int_{\rb}^{\rho(x,0)} \dd r
    \left[ \rho(x,0) - r \right] \frac{2 D(r)}{\sigma(r)}
    \:,
\end{equation}
and the functional $Q_T$ is the integrated current associated to the time evolution $\rho(x,t)$, obtained by comparing the number of particles to the right of the origin at final time $t=T$ and initial time $t=0$,
\begin{equation}
    Q_T[\rho] = \int_0^\infty \left[
    \rho(x,T) - \rho(x,0)
    \right] \dd x
    \:.
\end{equation}
For large $T$, the functional integrals can be computed by minimizing the action. Denoting $(q,p)$ the optimal values of $(\rho,H)$, this gives the MFT equations~\cite{Derrida:2009aSM},
\begin{align}
  \label{eq:MFT_qSM}
  \partial_t q &= \partial_x(D(q) q) - \partial_x[\sigma(q)\partial_x p]
  \:,
  \\
  \label{eq:MFT_pSM}
  \partial_t p &= - D(q) \partial_x^2 p - \frac{\sigma'(q)}{2}(\partial_x p)^2
  \:,
\end{align}
and the initial and final conditions,
\begin{equation}
    \label{eq:InitFinSM}
    p(x,T) = \lambda \Theta(x)
    \:,
    \quad
    p(x,0) = \lambda \Theta(x) +  \int_{\rb}^{q(x,0)} \frac{2 D(r)}{\sigma(r)} \dd r
    \:,
\end{equation}
where $\Theta$ is the Heaviside step function.

\bigskip

In the case of a semi-infinite system, connected to a reservoir at density $\rl$, the same procedure can be applied, and yields the same equations~(\ref{eq:MFT_qSM}-\ref{eq:InitFinSM}) but restricted to the positive axis $x>0$, and completed by the boundary conditions at the origin~\cite{Saha:2023SM},
\begin{equation}
    \label{eq:BoundCondMFTSM}
    q(0,t) = \rl
    \:,
    \quad
    p(0,t) = 0
    \:.
\end{equation}
From the solution of the MFT equations~(\ref{eq:MFT_qSM}-\ref{eq:BoundCondMFTSM}), the CGF~\eqref{eq:MGFfromMFTSM} can be computed as
\begin{equation}
    \ln \moy{\e^{\lambda Q_T}}
    \underset{T \to \infty}{\simeq} 
    \lambda Q_T[q] - S[q,p] - F[q(x,0)]
    \:.
\end{equation}
Rescaling $x$ by $\sqrt{T}$ and $t$ by $T$ in the action, one can show that $\ln \moy{\e^{\lambda Q_T}} \propto \sqrt{T}$~\cite{Derrida:2009aSM}. Similarly, the correlation profiles can be computed as
\begin{equation}
    \label{eq:ProffromMFT0SM}
    \frac{\moy{\eta_r \e^{\lambda Q_T}}}{ \moy{\e^{\lambda Q_T}} }
    =
    \frac{\displaystyle \int \mathcal{D} \rho(x,t) \mathcal{D} H(x,t)
    \int \mathcal{D} \rho(x,0) \: \rho(r,T)
    \e^{\lambda Q_T[\rho] - S[\rho,H] - F[\rho(x,0)]}
    }
    {
    \displaystyle \int \mathcal{D} \rho(x,t) \mathcal{D} H(x,t)
    \int \mathcal{D} \rho(x,0) \:
    \e^{\lambda Q_T[\rho] - S[\rho,H] - F[\rho(x,0)]}
    }
    \underset{T \to \infty}{\simeq} q(r,T)
    \:,
\end{equation}
which can be similarly rescaled by $\sqrt{T}$ in space and $T$ in time to yield~\cite{Poncet:2021SM,Grabsch:2022SM,Grabsch:2023SM},
\begin{equation}
    \label{eq:ProffromMFTSM}
    \frac{\moy{\eta_r \e^{\lambda Q_T}}}{ \moy{\e^{\lambda Q_T}} }
    \underset{T \to \infty}{\simeq} 
    q \left( x = \frac{r}{\sqrt{T}}, 1 \right)
    = \Phi(x)
    \:.
\end{equation}
These arguments justify the scaling forms~\eqref{eq:ScalingFromSM} introduced above.
The aim is thus to solve the MFT Eqs.~(\ref{eq:MFT_qSM}-\ref{eq:BoundCondMFTSM}) for $x>0$.

\subsection{Semi-infinite vs infinite geometry}

The MFT equations for the SEP are difficult to solve explicitly. A solution has recently been obtained in the infinite case using the inverse scattering technique~\cite{Mallick:2022SM}. It is a powerful method, but it is not clear how to apply it in the half-infinite case, in particular because it uses a mapping which involves the derivatives of both $q$ and $p$~\cite{Mallick:2022SM}, but we only have boundary conditions of the values of $q$ and $p$ at the origin~\eqref{eq:BoundCondMFTSM}, not their derivatives. Therefore, we will rely on a perturbative expansion in $\lambda$ to compute the first few orders of the solution of the MFT equations~(\ref{eq:MFT_qSM}-\ref{eq:BoundCondMFTSM}).

This is however still a difficult task, see for instance Refs.~\cite{Krapivsky:2015aSM,Grabsch:2022SM,Grabsch:2023SM,Grabsch:2024aSM} for related problems in the infinite geometry. Nevertheless, the solution of the MFT equations at final time, which thus gives the correlation profile~\eqref{eq:ProffromMFTSM}, in the infinite geometry has been shown to obey a simple closed integral equation~\cite{Grabsch:2022SM,Grabsch:2023SM,Mallick:2022SM}, which can be solved explicitly. The idea in this section is thus to express the solution $(q,p)$ of the MFT equations~(\ref{eq:MFT_qSM}-\ref{eq:BoundCondMFTSM}) for $x>0$ in the half-infinite geometry, in terms of the solution of the MFT equations~(\ref{eq:MFT_qSM}-\ref{eq:InitFinSM}) in the infinite geometry $x \in \mathbb{R}$, which we denote $(\qf, \pf)$, order by order. We denote
\begin{equation}
    q = \sum_{n \geq 0} \lambda^n q_n
    \:,
    \quad
    \qf = \sum_{n \geq 0} \lambda^n \qf_n
    \:,
    \quad
    p = \sum_{n \geq 1} \lambda^n p_n
    \:,
    \quad
    \pf = \sum_{n \geq 1} \lambda^n \pf_n
    \:.
\end{equation}

\subsection{Order 0}

At order $0$ in $\lambda$, the MFT equations for the SEP in the semi-infinite geometry~(\ref{eq:MFT_qSM}-\ref{eq:BoundCondMFTSM}) yield
\begin{equation}
    \partial_t q_0 = \partial_x^2 q_0
    \:,
    \quad
    q_0(x,0) = \rb
    \:,
    \quad
    q_0(0,t) = \rl
    \:.
\end{equation}
These equations can be solved explicitly, and yield,
\begin{equation}
    \label{eq:SolqOrder0}
    q_0(x,t) = \rb \: \erf \left( \frac{x}{2 \sqrt{t}} \right)
    + \rl \erfc \left(  \frac{x}{2 \sqrt{t}} \right)
    \:.
\end{equation}

This solution can be expressed in terms of the solution of the MFT equations in full space~(\ref{eq:MFT_qSM}-\ref{eq:InitFinSM}), with an initial step of density
\begin{equation}
    \partial_t \qf_0 = \partial_x^2 \qf_0
    \:,
    \quad
    \qf_0(x,0) = \rho_- \Theta(-x) + \rho_+ \Theta(x)
    \:.
\end{equation}
Indeed the solution takes the form,
\begin{equation}
    \qf_0(x,t) = \frac{\rho_-}{2} \erfc \left( \frac{x}{2 \sqrt{t}} \right)
    + \frac{\rho_+}{2} \erfc \left( -\frac{x}{2 \sqrt{t}} \right)
    \:,
\end{equation}
which is identical to~\eqref{eq:SolqOrder0}, provided we choose the densities $\rho_+ = \rb$ and $\rho_- = 2\rl - \rb$. Note that the density $\rho_-$ can be negative, so it does not correspond to a physical value, but nevertheless the MFT equations still admit a solution, which has a physical meaning for $x>0$ only. With this choice of $\rho_\pm$, we have the relation
\begin{equation}
    \label{eq:Relq0SM}
    q_0(x,t) = \qf_0(x,t)
    \quad \text{for} \quad
    x > 0
    \:.
\end{equation}
The idea is to proceed similarly at each order, without solving explicitly the equations.

\subsection{Order 1}

We proceed similarly at order 1 in $\lambda$. The MFT equations for $p_1$ and $\pf_1$ read,
\begin{equation}
    \partial_t p_1 = - \partial_x^2 p_1
    \:,
    \quad
    p_1(x,1) = 1
    \:,
    \quad
    p_1(0,t) = 0
    \:,
    \quad \text{for} \quad
    x>0
    \:,
\end{equation}
\begin{equation}
    \partial_t \pf_1 = - \partial_x^2 \pf_1
    \:,
    \quad
    \pf_1(x,1) = \Theta(x)
    \:,
    \quad \text{for} \quad
    x \in \mathbb{R}
    \:.
\end{equation}
The solutions can again be written explicitly as
\begin{equation}
    p_1(x,t) = \erf \left( \frac{x}{2 \sqrt{1-t}} \right)
    \:,
    \quad
    \pf_1(x,t) = \frac{1}{2} \erfc \left( -\frac{x}{2 \sqrt{1-t}} \right)
    \:.
\end{equation}
We straightforwardly find the relation
\begin{equation}
    \label{eq:Relp1SM}
    p_1(x,t) = 2 \pf_1(x,t) - 1
    \quad \text{for} \quad x>0
    \:.
\end{equation}

\bigskip

Concerning $q_1$ and $\qf_1$, the MFT equations read,
\begin{equation}
    \label{eq:MFTq1SM}
    \partial_t q_1 = \partial_x^2 q_1
    - 2 \partial_x[q_0(1-q_0) \partial_x p_1]
    \:,
    \quad
    q_1(x,0) = \rb(1-\rb) [p_1(x,0) - 1]
    \:,
    \quad
    q_1(0,t) = 0
    \quad
    \text{for} \quad x >0
    \:,
\end{equation}
\begin{equation}
    \label{eq:MFTq1FSM}
    \partial_t \qf_1 = \partial_x^2 \qf_1
    - 2 \partial_x[\qf_0(1-\qf_0) \partial_x \pf_1]
    \:,
    \quad
    \qf_1(x,0) = \qf_0(x,0)(1-\qf_0(x,0)) [\pf_1(x,0) - \Theta(x)]
    \:,
    \quad
    \text{for} \quad x \in \mathbb{R}
    \:.
\end{equation}
These equations cannot be solved at arbitrary time, so this is where relating the two problems is useful. Indeed, using~(\ref{eq:Relq0SM},\ref{eq:Relp1SM}) and the equations~(\ref{eq:MFTq1SM},\ref{eq:MFTq1FSM}), we easily check that
\begin{equation}
    q_1(x,t) = 2 \qf_1(x,t) + \Delta q_1(x,t)
\end{equation}
is solution of
\begin{equation}
    \label{eq:EqForDeltaq1SM}
    \partial_t \Delta q_1 = \partial_x^2 \Delta q_1
    \:,
    \quad
    \Delta q_1(x,0) = 0
    \:,
    \quad
    q_1(0,t) = - 2 \qf_1(0,t)
    \quad
    \text{for} \quad
    x>0
    \:.
\end{equation}
We have thus reduced the problem of solving the MFT equations~\eqref{eq:MFTq1SM} at order 1 to computing $\qf_1(0,t)$. This is a much simpler task, since Eq.~\eqref{eq:MFTq1FSM} is a diffusion equation with a source term,
\begin{equation}
    \qf_1(x,t) = -\int_0^t \dd t' \int_{-\infty}^\infty \dd x' \:
    \frac{\e^{-\frac{(x-x')^2}{4(t-t')}}}{2\sqrt{\pi (t-t')}}
    \partial_{x'}[\qf_0(x',t')(1-\qf_0(x',t')) \partial_{x'} \pf_1(x',t')]
    + \int_{-\infty}^\infty \dd x'  \frac{\e^{-\frac{(x-x')^2}{4t}}}{2\sqrt{\pi t}} \qf_1(x',0)
    \:.
\end{equation}
The first integral cannot be computed explicitly for all $x$, but it can be computed for $x=0$ using~\cite{Owen:1980SM} and yields
\begin{equation}
    \qf_1(0,t) = -\frac{\rho_+ - \rho_-}{4} (1 - \rho_+ - \rho_-)
    = -\frac{\rb - \rl}{2} (1-2\rl)
    \:.
\end{equation}
Since this value does not depend on time, the solution of~\eqref{eq:EqForDeltaq1SM} takes the simple form,
\begin{equation}
    \Delta q_1(x,t) = (\rb-\rl)(1-2\rl) \erfc \left( \frac{x}{2\sqrt{t}} \right)
    \:.
\end{equation}
Therefore, the solution $q_1(x,1)$ at final time reads
\begin{equation}
    \label{eq:Relq1SM}
    q_1(x,1) = 2 \qf_1(x,1) +  (\rb-\rl)(1-2\rl) \erfc \left( \frac{x}{2} \right)
    \:.
\end{equation}

\subsection{Order 2}

Proceeding similarly at order 2, we make the change of functions
\begin{equation}
    p_2(x,t) = 4 \pf_2(x,t) + \Delta p_2(x,t)
    \:.
\end{equation}
We obtain that $\Delta p_2$ obeys the diffusion equation,
\begin{equation}
    \partial_t \Delta p_2 = - \partial_x^2 \Delta p_2
    \:,
    \quad
    \Delta p_2(x,1) = 0
    \:,
    \quad
    \Delta p_2(0,t) - 4 \pf_2(0,t)
    \:.
\end{equation}
The value of $\pf_2(0,t)$ can be computed explicitly, and reads,
\begin{equation}
    \pf_2(0,t) = \frac{1}{8}(1-2 \rl)
    \:.
\end{equation}
Therefore,
\begin{equation}
    p_2(x,t) = 4 \pf_2(x,t) - \frac{1-2\rl}{2} \erfc \left( \frac{x}{2 \sqrt{1-t}} \right)
    \:.
\end{equation}

\bigskip

For $q_2$, we make the change of functions,
\begin{equation}
    \label{eq:ChFctq2SM}
    q_2 = 4 \qf_2 - 3 (1-2\rl) \qf_1 + 2 (1-2\rb)(1-2\rl) \qf_0 \pf_1 + 2 \rb (1-2\rl) \pf_1 + \Delta q_2
    \:,
\end{equation}
from which we deduce that $\Delta q_2$ obeys
\begin{equation}
    \partial_t \Delta q_2 = \partial_x^2 \Delta q_2
    \:,
    \quad
    \Delta q_2(x,0) = -4 \rb (1-\rb) (1-2\rl)
    \:,
\end{equation}
\begin{equation}
    \Delta q_2(0,t) = - 4 \qf_2(0,t)
    + 3 (1-2\rl) \qf_1(0,t)
    - 2 (1-2\rb)(1-2\rl) \qf_0(0,t) \pf_1(0,t)
    - 2 \rb (1-2\rl) \pf_1(0,t)
    \:.
\end{equation}
We have already computed the values at $x=0$ of all these functions, except $\qf_2$, which can be computed in a similar manner. We get,
\begin{equation}
    \qf_2(0,t) = \frac{1}{4}(1 - 2\rl) [2 \rb^2 - 4 \rb \rl - \rl (1-3 \rl)]
    \:.
\end{equation}
Therefore, we deduce
\begin{equation}
    \Delta q_2(x,t) = -4 \rb (1-\rb) (1-2\rl) \erf \left( \frac{x}{2 \sqrt{t}} \right)
    + \frac{1}{2} (1 - 2\rl) [ 4 \rb^2 - \rl + \rb (2 \rl - 5) ] 
    \erfc \left( \frac{x}{2 \sqrt{t}} \right)
    \:.
\end{equation}
Finally, from~\eqref{eq:ChFctq2SM}, we obtain at $t=1$,
\begin{multline}
    \label{eq:Relq2SM}
    q_2(x,1) = 4 \qf_2(x,1) 
    - 3 (1-2\rl) \qf_1(x,1) 
    + 2 (1-2\rb)(1-2\rl) \qf_0(x,1) + 2 \rb (1-2\rl)
    \\
     -4 \rb (1-\rb) (1-2\rl) \erf \left( \frac{x}{2} \right)
    + \frac{1}{2} (1 - 2\rl) [ 4 \rb^2 - \rl + \rb (2 \rl - 5) ] 
    \erfc \left( \frac{x}{2} \right)
    \:.
\end{multline}

\subsection{Summary of the MFT results}
\label{sec:SummaryMFTres}

We have expressed, at each order in $\lambda$, the solution of the MFT equations in the semi-infinite system, in terms of those in the infinite geometry. We can thus relate the correlation profiles in each geometry thanks to~\eqref{eq:ProffromMFTSM},
\begin{equation}
    \Phi(x) = q(x,1) = \sum_{n \geq 0} \lambda^n \Phi_n(x)
    \:,
    \quad
    \Phi^{(F)}(x) = \qf(x,1) = \sum_{n \geq 0} \lambda^n \Phi_n^{(F)}(x)
    \:.
\end{equation}
The expressions of $\Phi_n^{(F)}$ can be obtained from~\cite{Grabsch:2022SM,Grabsch:2023SM}. Actually, their derivatives take a simpler form,
\begin{equation}
    \partial_x \Phi_0^{(F)}(x) = \frac{\rho_+ - \rho_-}{2\sqrt{\pi}} \e^{- \frac{x^2}{4}}
    \:,
\end{equation}
\begin{equation}
    \partial_x \Phi_1^{(F)}(x) = 
    \frac{2 (\rho_-^2 - \rho_+^2) - 4 \rho_- (1-\rho_+)}{4\sqrt{\pi}} \e^{- \frac{x^2}{4}}
    + \frac{(\rho_+ - \rho_-)^2}{2 \sqrt{2 \pi}} \e^{- \frac{x^2}{2}} \erfc \left( \frac{x}{\sqrt{2}}\right)
    \:,
\end{equation}
\begin{multline}
    \partial_x \Phi_2^{(F)}(x) = 
    \left(
    -3 \rho_-^3-3 \rho_-^2 (\rho_+-2)-\rho_- (\rho_+-2)^2-(\rho_+-2) \rho_+^2
    \right)
    \frac{1}{16 \sqrt{\pi}} \e^{- \frac{x^2}{4}}
    \\
    + \left(
    \rho_-^3+3 \rho_-^2 (\rho_+-1)+\rho_- (2-5 \rho_+)
   \rho_+ +\rho_+^2 (\rho_+ +1)
    \right)
    \frac{\e^{-\frac{x^2}{8}}}{8 \sqrt{2\pi}} \erfc \left( \frac{x}{2 \sqrt{2}} \right)
    \\
    - (\rho_+ - \rho_-)^3 \frac{\e^{-\frac{x^2}{12}}}{8 \sqrt{3 \pi}}
     \left[
        \erfc \left( \frac{x}{2\sqrt{6}} \right) + \erfc \left( \frac{x}{\sqrt{6}} \right) 
        -4 \: \mathrm{T} \left( \frac{x}{2\sqrt{3}}, \sqrt{3} \right)
     \right]
    \:,
\end{multline}
where
\begin{equation}
  \label{eq:defOwenT}
  T(h,a) = \frac{1}{2\pi}
  \int_0^a \frac{\e^{-\frac{h^2}{2}(1+x^2)}}{1+x^2} \dd x
\end{equation}
is the Owen $T$ function~\cite{Owen:1980SM}.

\bigskip

These expressions, combined with the relations~(\ref{eq:Relq0SM},\ref{eq:Relq1SM},\ref{eq:Relq2SM}) give,
\begin{equation}
    \label{eq:dPhi0}
    \Phi_0'(x) = \frac{\rb - \rl}{\sqrt{\pi}} \e^{- \frac{x^2}{4}}
    \:,
\end{equation}
\begin{equation}
    \label{eq:dPhi1}
    \Phi_1'(x) = 
    -(\rb^2 + \rb (1-2 \rl))\frac{\e^{- \frac{x^2}{4}}}{\sqrt{\pi}}
    + (\rb-\rl)^2 \sqrt{\frac{2}{\pi}}
    \erfc \left( \frac{x}{2 \sqrt{2}} \right)
    \:,
\end{equation}
\begin{multline}
    \label{eq:dPhi2}
    \Phi_2'(x) = 
    \left( \rb^2 (1-2\rl) - \rl (1-2\rb) \right)
    \frac{\e^{- \frac{x^2}{4}}}{2 \sqrt{\pi}}
    - \left( 4 \rb^3 + \rb^2 (1-14 \rl) - \rl^3 (3+2\rl) + 2\rb \rl (1+6 \rl) \right)
    \frac{\e^{-\frac{x^2}{8}}}{\sqrt{2 \pi}} \erfc \left( \frac{x}{2 \sqrt{2}} \right)
    \\
    + (\rb - \rl)^3 \frac{4 \e^{-\frac{x^2}{12}}}{\sqrt{3 \pi}}
     \left[
        \erfc \left( \frac{x}{2\sqrt{6}} \right) + \erfc \left( \frac{x}{\sqrt{6}} \right) 
        -4 \: \mathrm{T} \left( \frac{x}{2\sqrt{3}}, \sqrt{3} \right)
     \right]
    \:.
\end{multline}

The expression of $\Phi(x)$ can be obtained from these expressions by integrating $\Phi'$ on $[x,+\infty[$, with the boundary condition $\Phi(+\infty) = \rb$. In particular, for $x=0$, we get the value $\Phi(0)$. We check that it coincides with~\eqref{eq:ValAtZeroSM}. Combined with the relation~\eqref{eq:BoundCondMicroSM}, this gives the CGF,
\begin{multline}
    \label{eq:CumulantsSM}
    \hat\psi = 2\lambda \frac{\rl - \rb}{\sqrt{\pi}}
    + \frac{\lambda^2}{\sqrt{\pi}}
    \left(
      -2 \left(\sqrt{2}-1\right) \rb^2+\left(4 \sqrt{2}-6\right) \rb
      \rl+\rb-2 \left(\sqrt{2}-1\right) \rl^2+\rl
    \right)
    \\
    + \frac{\lambda^3}{9 \sqrt{\pi}} (\rb-\rl) \left(
      3 (6 \rb (2 \rl-1)-6 \rl-1)
      + 18 \sqrt{2} \left(2 \rb^2-6 \rb \rl+\rb+2
        \rl^2+\rl\right)
      -32 \sqrt{3} (\rb-\rl)^2
    \right)
    + \mathcal{O}(\lambda^4)
    \:.
\end{multline}
This expression coincides with the first three cumulants given in~\cite{Saha:2023SM}.

\section{Derivation of the integral equation for the correlations}

\subsection{Inferring the equation from the first orders}

In the infinite geometry, the knowledge of the first orders of $\Phi$ allowed to infer the general structure of the correlation functions. Indeed, it was shown in~\cite{Grabsch:2022SM,Grabsch:2023SM} that the rescaled derivatives of $\Phi^{(F)}$
\begin{equation}
    \Omega_\pm^{(F)}(x) \equiv \hat\psi^{(F)} \frac{\partial_x \Phi^{(F)}(x)}{\partial_x \Phi^{(F)}(0^\pm)}
    \:,
    \quad \text{for} \quad
    x \gtrless 0
    \:,
\end{equation}
satisfy, up to order $3$ in $\lambda$, the closed integral equations,
\begin{equation}
  \label{eq:CoupledWHeqsSM}
    \Omega_\pm^{(F)}(x) + \int_0^\infty \Omega^{(F)}_\mp(\mp z) \Omega^{(F)}_\pm(x\pm z) \dd z
    = K^{(F)}(x)
      \:,
      \quad
      K^{(F)}(x) = \omega^{(F)} \frac{\e^{- \frac{x^2}{4}}}{2 \sqrt{\pi}}
      \:,
\end{equation}
with $\hat\psi^{(F)}$ the CGF of the current in the infinite geometry, and
\begin{equation}
    \label{eq:DefomegaFSM}
  \omega^{(F)} = \rho_- (\e^{\lambda} - 1) + \rho_+ (\e^{-\lambda}-1) + \rho_+ \rho_- (\e^{\lambda}-1)(\e^{-\lambda}-1)
  \:.
\end{equation}
In Refs.~\cite{Grabsch:2022SM,Grabsch:2023SM}, it was then argued that this equation actually holds at any order in $\lambda$, and this was later proved in~\cite{Mallick:2022SM}. Our goal is to follow the same approach, using the MFT results given in Section~\ref{sec:SummaryMFTres}.

\bigskip

Inspired by the infinite geometry, we define
\begin{equation}
    \label{eq:DefOmegaSM}
    \Omega(x) = \hat\psi \frac{\Phi'(x)}{\Phi'(0)}
    \:.
\end{equation}
We now look for an equation similar to~\eqref{eq:CoupledWHeqsSM}, but with  now only one function $\Omega$ since there is no physical domain $x<0$. We therefore adapt the l.h.s. of~\eqref{eq:CoupledWHeqsSM} (left), and compute
\begin{equation}
   \Omega(x) + \int_0^\infty \Omega(z) \Omega(x+z) \dd z
    \:,
\end{equation}
using the expressions of Section~\ref{sec:SummaryMFTres}, with the goal of identifying a general structure from the first orders in $\lambda$ computed from MFT. Using the expressions of Section~\ref{sec:SummaryMFTres} with the definition~\eqref{eq:DefOmegaSM}, we find that
\begin{multline}
    \Omega(x) + \int_0^\infty \Omega(z) \Omega(x+z) \dd z =
     \lambda  (\rl-\rb) \frac{\e^{- \frac{x^2}{4}}}{2 \sqrt{\pi}}
     + 2 \lambda ^2 \left(2 \rb^2-6 \rb \rl + \rb +2 \rl^2 + \rl\right)
     \frac{\e^{- \frac{x^2}{4}}}{2 \sqrt{\pi}}
     \\
     + \frac{2 \lambda ^3}{3} (\rb-\rl) (6 \rb (2 \rl-1)-6 \rl-1)
  \frac{\e^{- \frac{x^2}{4}}}{2 \sqrt{\pi}}
  + \mathcal{O}(\lambda^4)
  \:.
\end{multline}
Remarkably, all the error functions and Owen-T functions present in the lowest orders of the correlation profiles have canceled out in the computation. At every order in $\lambda$, there only remains a Gaussian term for the $x$ dependence. We infer that this striking fact remains true at all orders in $\lambda$, so that
\begin{equation}
    \label{eq:ClosedIntegEqSM}
    \Omega(x) + \int_0^\infty \Omega(z) \Omega(x+z) \dd z = 
    K(x)
    \:,
    \quad
    K(x) \equiv
    \gamma \frac{\e^{- \frac{x^2}{4}}}{2 \sqrt{\pi}}
    \:,
\end{equation}
with a parameter $\gamma$ to be determined, given at first orders by
\begin{equation}
    \label{eq:ExpGammaLowOrders}
  \gamma = 4 \lambda  (\rl-\rb)
  + 2 \lambda ^2 \left(2 \rb^2-6 \rb \rl + \rb +2 \rl^2 + \rl\right)
  \\
  +\frac{2 \lambda ^3}{3} (\rb-\rl) (6 \rb (2 \rl-1)-6 \rl-1)+
  \mathcal{O}\left(\lambda ^4\right)
  \:.
\end{equation}
Furthermore, the cumulant generating function is expected to be a function of the single parameter~\cite{Saha:2023SM}
\begin{equation}
    \label{eq:DefomegaSM}
  \omega = \rl (\e^{\lambda} - 1) + \rb (\e^{-\lambda}-1) + \rb \rl (\e^{\lambda}-1)(\e^{-\lambda}-1)
\end{equation}
only, which is the equivalent of $\omega^{(F)}$~\eqref{eq:DefomegaFSM} in the case of a semi-infinite geometry. We thus look for a simple function of $\omega$ which yields~\eqref{eq:ExpGammaLowOrders} at first orders. We find that
\begin{equation}
    \label{eq:ExprGammaSM}
    \gamma = 4 \omega (1+\omega)
\end{equation}
is the simplest function of $\omega$ that reproduces~\eqref{eq:ExpGammaLowOrders}. We check below that this value of $\gamma$ is consistent with the boundary conditions~(\ref{eq:ValAtZeroSM},\ref{eq:BoundCondMicroSM}) derived above from microscopic considerations.

\bigskip

Due to the similarities with the infinite case, we infer, as in Refs.~\cite{Grabsch:2022SM,Grabsch:2023SM}, that the closed equation~\eqref{eq:ClosedIntegEqSM} is exact at all orders in $\lambda$. We give below further arguments supporting this claim. But first, we show how Eq.~\eqref{eq:ClosedIntegEqSM} can be solved.

\subsection{Solution of the equation, and cumulants}

The main equation~\eqref{eq:ClosedIntegEqSM} can actually be mapped onto the same equation as in the infinite geometry~\eqref{eq:CoupledWHeqsSM}. Indeed, since the solution $\Omega_\pm^{(F)}$ of~\eqref{eq:CoupledWHeqsSM} is symmetric, we can define
\begin{equation}
    \tilde\Omega^{(F)}(x) = \Omega_+^{(F)}(x) = \Omega_-^{(F)}(-x)
    \:.
\end{equation}
Rewriting now~\eqref{eq:CoupledWHeqsSM} in terms of $\tilde\Omega^{(F)}$, we get the exact same equation as in the semi-infinite geometry,
\begin{equation}
    \tilde\Omega^{(F)}(x) + \int_0^\infty \tilde\Omega^{(F)}(z) \tilde\Omega^{(F)}(x+z) \dd z = K^{(F)}(x)
    \:,
\end{equation}
but with a different kernel $K^{(F)}$. We can thus straightforwardly use the solution of~\eqref{eq:CoupledWHeqsSM} given in~\cite{Grabsch:2022SM,Grabsch:2023SM}, which thus gives,
\begin{equation}
  \label{eq:SolIntegEqSM}
  \int_0^\infty \Omega(x) \e^{\I k x} \dd x
  =    \exp \left[
    \frac{1}{2\pi} \int_0^\infty \dd x \: \e^{\I k x}
    \int_{-\infty}^{+\infty} \dd u \: \e^{-\I u x} \:
    \ln (1+\hat{K}(u))
  \right] - 1
  \:,
\end{equation}
where the Fourier transform of $K$ is defined as
\begin{equation}
    \hat{K}(k) = \int_{-\infty}^\infty K(x) \e^{\I k x} \dd x
    = \gamma \: \e^{-k^2}
    \:.
\end{equation}

In particular, setting $k = \I s$ and letting $s \to 0$, we obtain,
\begin{equation}
    \Omega(0) = \frac{1}{2\pi} \int_{-\infty}^\infty \ln (1+ \hat{K}(k)) \dd k
    = - \frac{1}{2\sqrt{\pi}} \mathrm{Li}_{\frac{3}{2}}(-\gamma)
    \:,
\end{equation}
where $\mathrm{Li}_s$ is the polylogarithm function. From the definition of $\Omega$~\eqref{eq:DefOmegaSM}, we get
\begin{equation}
    \label{eq:CGFSM}
    \hat\psi = \Omega(0) = - \frac{1}{2\sqrt{\pi}} \mathrm{Li}_{\frac{3}{2}}(-\gamma)
    \:,
\end{equation}
as announced in the main text.

\subsection{A consistency check}

We can actually check that the integral equation~\eqref{eq:ClosedIntegEqSM}, together with the boundary conditions~(\ref{eq:ValAtZeroSM},\ref{eq:BoundCondMicroSM}) is consistent with the expression of $\gamma$~\eqref{eq:ExprGammaSM}. For this, we use the solution~\eqref{eq:SolIntegEqSM} at $k=0$, which gives
\begin{equation}
  \int_0^\infty \Omega(x) \dd x
  =  \exp \left[- \sum_{n=1}^\infty \frac{(-\gamma)^n}{2 n} \right] - 1
  = \sqrt{1 + \gamma} - 1
  \:.
\end{equation}
Combined with the definition of $\Omega$~\eqref{eq:DefOmegaSM}, this gives,
\begin{equation}
    \hat\psi \frac{\rb - \Phi(0)}{\Phi'(0)} = \sqrt{1 + \gamma} - 1
    \:.
\end{equation}
Using now the boundary conditions obtained from the microscopic calculations~(\ref{eq:ValAtZeroSM},\ref{eq:BoundCondMicroSM}), and solving the above equation for $\gamma$, we obtain exactly the expression~\eqref{eq:ExprGammaSM}.

\subsection{Numerical validation}

\begin{figure}
    \centering
    \includegraphics[width=0.8\textwidth]{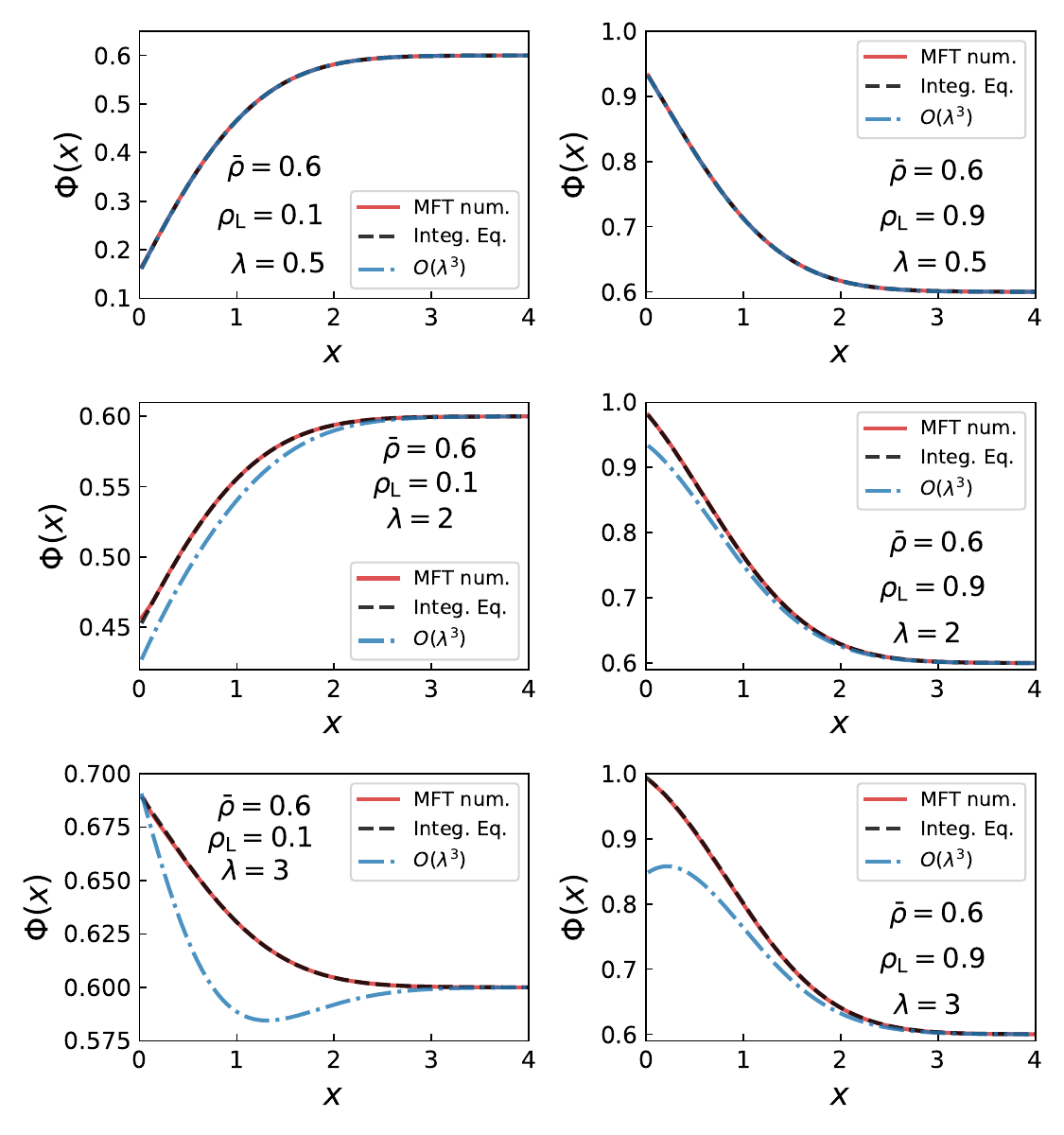}
    \caption{Numerical solution of the MFT equations~(\ref{eq:MFT_qSM}-\ref{eq:BoundCondMFTSM}) (solid red line), compared to the solution of the integral equation~\eqref{eq:ClosedIntegEqSM}, together with the boundary conditions~(\ref{eq:ValAtZeroSM},\ref{eq:BoundCondMicroSM}) (dashed black line) and the solution up to order $3$ in $\lambda$ given by~(\ref{eq:dPhi0}-\ref{eq:dPhi2}) (dotted blue line). Left: for $\rb = 0.6$, $\rl=0.1$. Right: for $\rb = 0.6$, $\rl=0.9$. Top: $\lambda = 0.5$. Middle: $\lambda = 2$. Bottom: $\lambda = 3$. }
    \label{fig:CompNum}
\end{figure}

To confirm the validity of the closed integral equation~\eqref{eq:ClosedIntegEqSM}, we compare the solution obtained for the profile $\Phi$ from~\eqref{eq:ClosedIntegEqSM} ---combined with the boundary conditions~(\ref{eq:ValAtZeroSM},\ref{eq:BoundCondMicroSM})--- to the numerical resolution of the MFT equations~(\ref{eq:MFT_qSM}-\ref{eq:BoundCondMFTSM}). To obtain this numerical solution, we use the algorithm described in Ref.~\cite{Krapivsky:2012SM} (this was for an infinite geometry, but it can be straightforwardly extended to the semi-infinite case). Note that to solve numerically Eqs.~(\ref{eq:MFT_qSM},\ref{eq:MFT_pSM}), we need to regularise the discontinuous boundary conditions~(\ref{eq:InitFinSM},\ref{eq:BoundCondMFTSM}) by a smooth function. This leads to small numerical errors near $x = 0$ in the numerical solution of the MFT equations.

The comparison is shown in Fig.~\ref{fig:CompNum} for different values of $\rl$, $\rb$. We also compare these results with the exact solution of the MFT equation computed up to order $3$ in $\lambda$, given in Section~\ref{sec:SummaryMFTres}. These plots show that all results are in perfect agreement for $\lambda = 0.5$. However, for $\lambda=2$ the solution truncated at $O(\lambda^3)$ deviates from the true MFT result, showing that higher orders in $\lambda$ are essential to describe the solution. This deviation becomes important for $\lambda=3$. For all values of $\lambda$, the result of our integral equation~\eqref{eq:ClosedIntegEqSM} is in perfect agreement with the numerical solution of the MFT equations, indicating that our results hold beyond the perturbative regime from which~\eqref{eq:ClosedIntegEqSM} was inferred.

\begin{figure}
    \centering
    \includegraphics[width=0.5\textwidth]{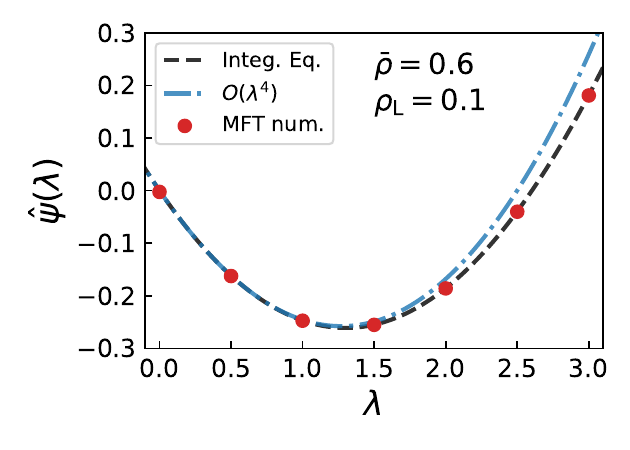}
    \caption{Cumulant generating function $\hat\psi(\lambda)$ as a function of $\lambda$, for $\rb = 0.6$ and $\rl = 0.1$. The red points are obtained from the numerical resolution of the MFT equations~(\ref{eq:MFT_qSM}-\ref{eq:BoundCondMFTSM}). The dashed black line corresponds to our result~\eqref{eq:CGFSM}, which results from the closed integral equation~\eqref{eq:ClosedIntegEqSM}. The dotted blue line corresponds to the first three orders in $\lambda$, given in~\eqref{eq:CumulantsSM}.}
    \label{fig:CompNumPsi}
\end{figure}

Finally, we also show in Fig.~\ref{fig:CompNumPsi} the cumulant generating function $\hat\psi$ as a function of $\lambda$, obtained from the numerical resolution of the MFT equations and compared to our result~\eqref{eq:CGFSM}. For $\lambda \gtrsim 1.5$, we observe a clear deviation between the MFT result and the perturbative result up to $O(\lambda^3)$~\eqref{eq:CumulantsSM}. In this regime, our prediction from the integral equation~\eqref{eq:ClosedIntegEqSM} remains in excellent agreement with the MFT result.

All these points further confirm the exactness of the integral equation~\eqref{eq:ClosedIntegEqSM}, for arbitrary value of $\lambda$.

\section{Applications}

We now present two applications of our results, which solve two problems that have remained open up to now.

\subsection{The survival probability of a fixed target}

The first application is the survival probability of a fixed target in the SEP. The survival probability is the probability $S(t)$ that no particle has touched the target up to time $t$. As usual in this context~\cite{Redner:2001SM}, the survival probability is computed by placing an absorbing wall at the position of the target. This absorbing wall is actually equivalent to a reservoir which cannot inject particles, i.e. with $\alpha = 0$. It thus corresponds to $\rl = 0$. The survival probability $S(t)$ therefore corresponds to the probability that no particle has entered the reservoir, i.e.,
\begin{equation}
    \label{eq:LinkSurvProbCurrSM}
    S(t) = \mathbb{P}(Q_t = 0)
    \:.
\end{equation}
The distribution of $Q_t$ can be obtained from the CGF~\eqref{eq:CGFSM} through an inverse Laplace transform, which at large times reduces to a Legendre transform,
\begin{equation}
  \mathbb{P}(Q_t = q \sqrt{t}) \underset{t \to \infty}{\simeq} \e^{- \sqrt{t} \phi(q)}
  \:,
  \quad
  \text{where}
  \quad
\phi(q) = - \hat\psi(\lambda^\star(q)) 
    - q \lambda^\star(q)
  \:,
  \quad \text{and} \quad
  \hat\psi'(\lambda^\star(q)) = q
  \:.
\end{equation}
Setting $q=0$, we obtain that the survival probability~\eqref{eq:LinkSurvProbCurrSM} reads
\begin{equation}
    S(t) \underset{t \to \infty}{\simeq}
    \e^{-\sqrt{t} \: F(\rb)}
    \:,
\end{equation}
where
\begin{equation}
    F(\rb) = - \hat\psi(\lambda^\star)
    \:,
    \quad
    \hat\psi'(\lambda^\star) = 0
    \:,
    \quad
    \text{with}
    \quad
    \rl = 0
    \:.
\end{equation}
Since, for $\rl = 0$, 
\begin{equation}
    \hat\psi(\lambda) = - \frac{1}{2\sqrt{\pi}}
    \mathrm{Li}_{\frac{3}{2}}
    \left[
    - 4 \rb (\e^{-\lambda}-1)(1+ \rb (\e^{-\lambda}-1) )
    \right]
\end{equation}
has a singularity for $\rb > \frac{1}{2}$ at $\lambda = \ln \frac{2 \rb}{2 \rb - 1}$, this procedure can only be carried out explicitly for $\rb \leq \frac{1}{2}$. Taking the derivative with respect to $\lambda$, we get,
\begin{equation}
    \hat\psi'(\lambda)
    =  \e^{-\lambda} \frac{1}{2\sqrt{\pi}}
    \frac{1 + 2\rb (\e^{-\lambda}-1)}{(\e^{-\lambda}-1) (1 + \rb (\e^{-\lambda}-1))}
    \mathrm{Li}_{\frac{1}{2}}
    \left[
    - 4 \rb (\e^{-\lambda}-1)(1+ \rb (\e^{-\lambda}-1) )
    \right]
    \:.
\end{equation}
The solution of $\hat\psi'(\lambda^\star) = 0$ is thus given by $\lambda^\star = +\infty$. And therefore we get
\begin{equation}
    S(t) \underset{t \to \infty}{\simeq}
    \e^{-\sqrt{t} \: F(\rb)}
    \:,
    \quad
    F(\rb) = \frac{1}{2 \sqrt{\pi}} \mathrm{Li}_{\frac{3}{2}} \left[ 4 \rb (1-\rb)\right]
    \:.
\end{equation}

\subsection{The SEP with a localised source}

The second application of our results concern the SEP with a localised source. This problem was introduced in~\cite{Krapivsky:2012aSM,Krapivsky:2014cSM} and consists in an infinite SEP, initially empty, coupled to a reservoir on site $0$ which can only inject particles. If the injection rate is sufficiently fast, the site $0$ is always occupied, so that this problem can be described in terms of two independent semi-infinite SEP, with initially $\rb = 0$, coupled to reservoirs at density $\rl = 1$. The number $N_t$ of particles injected up to time $t$ is therefore the sum of the current $Q_t$ injected in the two half-infinite systems. Since they are independent, we get,
\begin{equation}
    \frac{1}{\sqrt{t}} \ln \moy{\e^{\lambda N_t}}
    = \frac{1}{\sqrt{t}} \ln \moy{\e^{\lambda Q_t}}^2
    \underset{t \to \infty}{\simeq} 2 \hat\psi(\lambda)
    = - \frac{1}{\sqrt{\pi}} \mathrm{Li}_{\frac{3}{2}} \left[-4 \e^{\lambda} (\e^{\lambda}-1) \right]
    \:,
    \quad
    \text{for}
    \quad
    \rl = 1 \:, \quad \rb = 0
    \:.
\end{equation}
Expanding in powers of $\lambda$, we check that we recover the first two cumulants computed in~\cite{Krapivsky:2012aSM},
\begin{equation}
    \frac{\moy{N_t}}{\sqrt{t}} \underset{t \to \infty}{\simeq}
    \frac{4}{\sqrt{\pi}}
    \:,
    \frac{\moy{N_t^2}_c}{\sqrt{t}} \underset{t \to \infty}{\simeq}
    \frac{4(3-2\sqrt{2})}{\sqrt{\pi}}
    \:.
\end{equation}

\end{document}